\documentclass[12pt,preprint]{aastex}
\usepackage{amssymb}
\usepackage{longtable}
\usepackage{amsmath}
\usepackage[normalem]{ulem} % strikethrough

\newcommand{\zdot}{\makebox[0pt][l]{.}} % arcmin with decimal dot
\newcommand{\up}[1]{\ifmmode^{\rm #1}\else$^{\rm #1}$\fi}

\newcommand{\arcd}{\ifmmode^{\circ}\else$^{\circ}$\fi}
\newcommand{\arcm}{\ifmmode{'}\else$'$\fi}
\newcommand{\arcs}{\ifmmode{''}\else$''$\fi}

\begin{document}
\title{Transformation of the equatorial proper motion to the galactic system}

\author{Rados\l{}aw Poleski}
\affil{Department of Astronomy, Ohio State University,
140 W.\ 18th Ave., Columbus, OH 43210, USA; 
poleski@astronomy.ohio-state.edu\\
Warsaw University Observatory, Al. Ujazdowskie 4, 00-478 Warszawa, Poland
}

%%%%%%%%%%%%%%%%%%%%%%%%%%%%%%%%%%%%%%%%%%%%%%%%%%%%%%%%%%%%%%%%%%%%%%%%%
%                               Abstract
%%%%%%%%%%%%%%%%%%%%%%%%%%%%%%%%%%%%%%%%%%%%%%%%%%%%%%%%%%%%%%%%%%%%%%%%%
\begin{abstract}

I present simple analytical equations to transform proper motion vectors from equatorial to Galactic coordinates. 

\end{abstract}


%%%%%%%%%%%%%%%%%%%%%%%%%%%%%%%%%%%%%%%%%%%%%%%%%%%%%%%%%%%%%%%%%%%%%%%%%
%                           Introduction
%%%%%%%%%%%%%%%%%%%%%%%%%%%%%%%%%%%%%%%%%%%%%%%%%%%%%%%%%%%%%%%%%%%%%%%%%
%\section{Introduction}

The proper motions of celestial bodies are typically measured in the equatorial system of coordinates ($\mu_{\alpha\star}=\mu_{\alpha}\cos\delta$ and $\mu_{\delta}$).
The great majority of objects for which proper motions give astrophysical information lie in the Galaxy.
Thus, to analyze these data one needs to transform the measured values to the Galactic coordinate system ($\mu_{l\star}=\mu_{l}\cos b$ and $\mu_{b}$). 
One way to do that is by representing the proper motion as a vector in a three-dimensional Cartesian system and applying a coordinate system rotation i.e.,\,by multiplying the vector by a matrix, and projecting the result onto the sphere. 
The conversion of the coordinates between the equatorial system ($\alpha$, $\delta$)
and the Galactic one ($l$, $b$) is described by the following equations \citep[e.g.,][]{binney98}:
%\begin{equation}
\begin{align}
%b = \arcsin{\left(
\sin b & = & 
\cos\delta\cos\delta_G\cos{\left(\alpha-\alpha_G\right)} + \sin\delta\sin\delta_G \label{eq:b}\\
%\right)}
%\label{equ:b}
%\end{equation}
%\begin{equation}
\sin(l_{NGP}-l)\cos b & = & \cos\delta\sin{\left(\alpha-\alpha_G\right)} \\
%l = \arctan\left(
%\frac{\sin\delta-\sin b\sin\delta_G}{\cos\delta\sin{\left(\alpha-\alpha_G\right)}\cos\delta_G}
%\right) + l_{NGP}
%\label{equ:l}
%\end{equation}
%\begin{equation}
\cos(l_{NGP}-l)\cos b & = & \sin\delta\cos\delta_G-\cos\delta\sin\delta_G\cos{\left(\alpha-\alpha_G\right)}
%\end{equation}
\end{align}
where $\alpha_G$ and $\delta_G$ are the equatorial coordinates of the North Galactic Pole 
and $l_{NGP}$ denotes the Galactic longitude of the North Celestial Pole. 
The values of these quantities are \citep{perryman97}:
\begin{equation}
\alpha_G = 192\zdot\arcd85948,\qquad 
%\end{equation}
%\begin{equation}
\delta_G = 27\zdot\arcd12825,\qquad
%\end{equation}
%\begin{equation}
l_{NGP} = 122\zdot\arcd93192
\end{equation}
The transformation of the proper motions can be represented as a rotation:
\begin{equation}
\left(\begin{array}{c}
\mu_{l\star} \\
\mu_b 
\end{array}\right)
 = 
\left( \begin{array}{cc}
A_{11} & A_{12} \\
A_{21} & A_{22}
\end{array} \right)
\left(\begin{array}{c}
\mu_{\alpha\star} \\
\mu_{\delta} 
\end{array}\right)
\end{equation}
where $A_{11} = \partial \mu_{l\star}/\partial \mu_{\alpha\star}$ % = \partial (l\cos b)/\partial (\alpha\cos\delta)$ 
etc. The values of $A_{21}$ and $A_{22}$ can be trivially found by differentiating the equation (\ref{eq:b}). 
At the same time, calculating $A_{11}$ and $A_{12}$ by the same method results in some tedious algebra. 
Instead of doing this, one can use the fact that the rotation matrix has to be orthogonal matrix with the determinant of $1$ i.e.,\, $A_{11}=A_{22}$ and $A_{12}=-A_{21}$. 
The resulting transformation is: 
\begin{equation}
\left(\begin{array}{c}
\mu_{l\star} \\
\mu_b 
\end{array}\right)
 = \frac{1}{\cos b}
\left( \begin{array}{cc}
C_1 & C_2 \\
-C_2 & C_1
\end{array} \right)
\left(\begin{array}{c}
\mu_{\alpha\star} \\
\mu_{\delta} 
\end{array}\right) \label{eq:fin}
\end{equation}
where the coefficients $C_1$, $C_2$ are given by:
\begin{eqnarray}
 C_1 & = & \sin\delta_G\cos\delta-\cos\delta_G\sin\delta\cos{\left(\alpha-\alpha_G\right)} \label{eq:C1}\\ 
 C_2 & = & \cos\delta_G\sin{\left(\alpha-\alpha_G\right)} \label{eq:C2}
\end{eqnarray}
The quantity $\cos b$ can be calculated as 
\begin{equation}
\cos b = \sqrt{C_1^2+C_2^2}
\end{equation}

The above equations define the way to transform proper motions from equatorial to Galactic coordinates even without direct calculation of the Galactic coordinates of the object. 
The equations for transformation of proper motions do not appear in many textbooks.
They are included in \citet{smart38} and were later reproduced by \citet{bovy11}.

\acknowledgments

I thank Andy Gould for discussion and Jo Bovy for pointing appropriate references.  
I also thank Nick Rowell for pointing out inaccuracy. 
This work was supported by Polish Ministry of Science and Higher Education through the program ,,Iuventus Plus'' award No. IP2011 043571.

%%%%%%%%%%%%%%%%%%%%%%%%%%%%%%%%%%%%%%%%%%%%%%%%%%%%%%%%%%%%%%%%%%%%%%%%%
%                           Bibliography
%%%%%%%%%%%%%%%%%%%%%%%%%%%%%%%%%%%%%%%%%%%%%%%%%%%%%%%%%%%%%%%%%%%%%%%%%

\end{document}